# Anomalous magnetic and transport properties of La$_{0.8-x}$Eu$_x$Sr$_{0.2}$MnO$_3$ (0.04 ≤ $x$ ≤ 0.36) compounds

[a,b]Rakesh Kumar, [a]C.V. Tomy, [b]P.L. Paulose, [c]R. Nagarajan and [b]S.K. Malik

[a]*Indian Institute of Technology, Bombay, Mumbai – 400 076, India*
[b]*Tata Institute of Fundamental Research, Colaba, Mumbai – 400 005, India*
[c]*EMCR, DAE, Mumbai – 400 005, India*

Anomalous magnetic and transport properties observed in the La$_{0.8-x}$Eu$_x$Sr$_{0.2}$MnO$_3$ (0.04 ≤ $x$ ≤ 0.36) compounds are presented in this paper. The Curie temperature ($T_C$) decreases from 260 K for $x$ = 0.04 to 188 K for $x$ = 0.16 and surprisingly thereafter increases for higher Eu concentrations ($x$ > 0.16) and becomes nearly constant ~230 K. Resistivity increases with Eu concentration up to $x$ = 0.16 but decreases for higher Eu concentration ($x$ > 0.16). In the magnetoresistance data, in addition to a peak corresponding to the insulator-metal transition at $T_C$ ($T_{I-M1}$), a second peak is also observed at a relatively lower temperature, ($T_{I-M2}$). Both $T_{I-M1}$ and $T_{I-M2}$ follow the same trend as $T_C$. The unique variation of $T_C$ and magnetoresistance is explained on the basis of structure and disorder correlated to microscopic electronic phase segregation. The studies with the existing studies also point to the universibility of this consideration.

*Corresponding author:* rakesh.kumar@impmc.jussieu.fr, Tel.: +33 0178092746, Fax: +33 0144273785

Manganites having the general formula $Ln_{1-x}AE_x MnO_3$ ($Ln$ = rare earth metal, $AE$ = alkaline earth metal) show a large drop in the resistance near the insulator-metal transition temperature ($T_{I-M}$) on application of a magnetic field [1-4]. These compounds are important not only for the possible applications such as memory devices, magnetic field sensors, etc., but also for the studies of strongly correlated electron systems. Due to the strong electron lattice correlation, these systems show interesting magnetic and transport properties on partial substitution of the alkaline earth (or rare earth) metal at the $A$-site or the transition metal (Mn) at the $B$-site. The magnetic ordering and transport properties of the compounds are decided by the $B$-site ions only. The doping concentration of charge carriers and the average $A$-site ionic radii decide the $Mn^{3+}/Mn^{4+}$ ratio and the hopping amplitude of the charge carriers, respectively. In the present work, we have fixed the $Mn^{3+}/Mn^{4+}$ ratio (by fixing the Sr concentration) and studied the effect of average ionic radii close to the ferromagnetic metallic (FM)/ ferromagnetic insulating (FI) phase boundary. We have selected the parent compound $La_{0.8}Sr_{0.2}MnO_3$ (Curie temperature $T_C$ = 309 K) close to the bi-critical region ($La_{0.83}Sr_{0.17}MnO_3$), interfacing the FM and FI phases [5]. We have substituted $Eu^{3+}$ ($4f^6$, $J = 0$) at the $La^{3+}$ site to avoid complications due to other magnetic interactions, since $Eu^{3+}$ ions are nonmagnetic in nature. Here we have tuned only the average ionic radii at the $A$-site and have discussed in this paper the anomalous magnetic and transport properties resulting from this tuning.

The compounds, $La_{0.8-x}Eu_xSr_{0.2}MnO_3$ ($x$ = 0.04, 0.08, 0.16, 0.24, 0.36), were prepared by the standard solid state reaction method using powders of $La_2O_3$, $Eu_2O_3$, $SrCO_3$ and MnO. The final sintering of the pelletized powder was carried out at 1250°C for twenty four hours. The low field and high field magnetization measurements were performed in a SQUID magnetometer (Quantum Design, USA) and Vibrating Sample Magnetometer (Oxford Instruments, UK), respectively. The transport properties were measured in a Physical Property Measurement System (Quantum Design, USA).

The x-ray diffraction (XRD) patterns of the compounds were analyzed by the Rietveld method using Fullprof software, which show them to form in a single phase orthorhombic structure (space group *Pnma*, No. 62). To check the homogeneity at microscopic level, the samples were examined by a scanning electron microscope coupled with EDX (Energy dissipative x-ray analyzer). The analyses showed the compounds to be homogeneous except for a few spots of $Mn_3O_4$ which is too small to be observed in XRD. The lattice parameters obtained from the Rietveld analyses is shown in Fig. 1. The lattice parameter *b* and *c* and the unit cell volume shows a monotonic decrease on increasing the Eu concentration. The decrease in volume is expected due to the relatively smaller ionic radius of $Eu^{3+}$ (1.12 Å) compared to that of $La^{3+}$ (1.216 Å) ions, which in turn indicates the incorporation of Eu ions to the lattice sites. However, the behaviour of lattice parameter *a* w.r.t. concentration does not follow this trend and seems to have a bearing on the property of the system (see latter). We also note that the substitution of Eu drives the parent compound from rhombohedral to orthorhombic structure, as it happens in $La_{1-x}Sr_xMnO_3$ system itself for *x* < 0.175 [5] and also as it happens in the case of other related rare earth (*Ln*) systems, such as $Ln_{0.7}AE_{0.3}MnO_3$ [6]

Figure 2 shows the results of the magnetization measurements of the $La_{0.8-x}Eu_xSr_{0.2}MnO_3$ (0.04 ≤ *x* ≤ 0.36) compounds. The $T_C$ for the parent compound is reported to be of 309 K [5], which decreases monotonically with Eu substitution to 188 K for *x* = 0.16 and then surprisingly increases and becomes nearly constant around 230 K for higher Eu concentrations (*x* ≥ 0.24). The $T_C$ variation as a function of Eu concentration *x* is shown in Fig. 3. To the best of our knowledge, this is the first instance where an increase in the $T_C$ is observed on dilution of the *A*-site. Both the $La^{3+}$ and $Eu^{3+}$ ions are nonmagnetic in nature and hence the Eu substitution at the La site is expected only to create cationic size disorder at the *A*-site, which is quantified by a parameter 'variance' [6, 7] defined as $\sigma^2 = \Sigma y_i r_i^2 - \langle r_A \rangle^2$, where, $y_i$ is the fractional occupancy, $r_i$ is the ionic radii, $\langle r_A \rangle$ and is the average ionic radii at A-site. The variance for the compounds $La_{0.8-x}Eu_xSr_{0.2}MnO_3$ increases monotonically as Eu concentration is increased (see Fig. 3), which indicates increased cationic disorder at the *A*-site. The weak magnetic anomaly observed at 45 K in the magnetization data is due to the ferrimagnetic transition of the minor impurity phase $Mn_3O_4$ [8, 9]. The magnetization measurements at 5 K (inset Fig. 2) for all the compounds show almost the same saturation magnetic moment of

~3.6 $\mu_B$/ Mn which is only slightly lower than the expected value (3.8 $\mu_B$ / Mn for this system). Similar values of the saturation moment confirms the same $Mn^{3+}/Mn^{4+}$ ratio for all the compounds of the series $La_{0.8-x}Eu_xSr_{0.2}MnO_3$ (0.04 ≤ $x$ ≤ 0.36). The slightly reduced value of the saturation moment may be due the canting of some of the spins, which is expected in these compounds [10].

Figure 4 shows the results of resistivity (left panel) and magnetoresistance (right panel) measurements on the compounds. In addition to the insulator-metal transition ($T_{I-M}$) at 272 K close to its $T_C$ of 260 K for the $x$ = 0.04 compound ($La_{0.8-x}Eu_xSr_{0.2}MnO_3$), an anomaly develops at 210 K in the resistivity data. For $x$ = 0.08, two peaks are observed, one at 236 K (close to its $T_C$ of 228 K) and the other at a lower temperature of 156 K. For higher Eu concentrations ($x$ > 0.08), no insulator-metal transition is observed corresponding to $T_C$ but only the second transition at temperatures far below the $T_C$. But all the compounds show two transitions in the magnetoresistance, one corresponding to the $T_C$ and the other corresponding to the second transition at lower temperatures. The high temperature peak (corresponding to $T_C$) is denoted as $T_{I-M1}$ and the low temperature peak as $T_{I-M2}$, respectively. The behaviour of $T_{I-M1}$ and $T_{I-M2}$ with respect to $x$ follows that of the $T_C$-(Fig. 3); which decreases initially and then increases for Eu concentrations $x$ > 0.16 and remain constant thereafter.

We interpret the origin of the second magnetoreistance peak (occurring at a lower temperature) as the result of two competing phases, both of ferromagnetic nature, since no anomaly is observed corresponding to the second transition in the dc magnetization (Fig. 2). The two competing phases are a ferromagnetic metallic (FM) phase and a ferromagnetic insulating (FI) phase. The FI phase arises in a background of FM phase, due to local strain arising from a change in Mn-O-Mn bond angle due to Eu substitution (as $Eu^{3+}$ has a smaller ionic radius compared to $La^{3+}$ - Valence state of Eu is established to be 3+ from Eu Mössbauer measurements [11]). It is well known in manganites that the bond angle controls the charge hopping and hence change in resistance on Eu substitution. In other words, the development of another phase is a result of the strain in the lattice due to the substitution of an ion with a smaller ionic radius. This consideration is substantiated by the fact that the features of the ρ(T) curve changes with the cross over of lattice parameter *a* and *c* (Fig. 1). For $x = 0.04$ ($a < c$), the FM phase dominates over the growing FI phase resulting only in a hump below $T_C$. As the Eu concentration is increased to $x = 0.08$ ($a < c$), the insulating phase grows but still the metallic phase dominates at low temperatures and so its competition with insulating phase gives an illusion of second insulator-metal transition below $T_C$. The FI phase due to strain becomes evident from the $x = 0.16$ (*a* is closer to *c*) compound where the insulating phase dominates over the ferromagnetic metallic phase and hence we do not see any insulator-metal transition corresponding to $T_C$. For higher Eu concentrations ($x > 0.16$; $a > c$), the strain in the system may be very large which forces the system to relax into two independent phases for the minimization of energy of the system, and hence a phase segregation occurs. In support of the above, we note that a similar situation, viz., feature of ρ(T) being dependent on whether $a < c$ or $c < a$ has been observed in $Ln_{0.7}AE_{0.3}MnO_3$ system [6]. Further, we point out that Akahoshi *et al* have suggested that a disorder near the bicritical region would produce a phase separation into two competing ordered phases [12]. The parent compound in our studies, $La_{0.8}Sr_{0.2}MnO_3$, is near to the interface of FM and FI phases [5]. Hence the *A*-site disorder produced by the substitution of Eu on the La site can result in a strain which result in the development of a FI phase in the background of a FM phase.

With regard to dependence of $T_C$ on *x*, investigations of Rodriguez-Martinez and Attfield [6] on $Ln_{0.7}AE_{0.3}MnO_3$ have shown that anomalies / behaviour of structure, conductivity and magnetism go together and depend on $\sigma^2$. This is the case in our system also, as seen from the variation of $T_C$, $T_{I-M1}$, and $T_{I-M2}$ with respect to *x* (Fig. 3). As in their case, in our case also the change from decreasing trend of a parameter to increasing trend occurs around the value of *x* where the cross over of *a* and *c* occurs. However, they have also pointed out that the $T_M$ is not a simple function of variance above a critical value of $\sigma^2$ and will become constant and independent of variance [6] due to phase segregation. Our results are the best verification of Rodriguez suggestion of properties being dependent on $\sigma^2$ and whether $a < c$ or $c < a$, in a single series of the compounds. (Results of Rodriguez and Attfield were with substitution of different elements).

In summary, we have shown that the *A*-site disorder in $La_{0.8-x}Eu_xSr_{0.2}MnO_3$ compounds leads to a homogeneous structural phase but microscopically dual electronic phase. The studies also confirm that the structural, conducting and magnetic properties depend on disorder in a systematic way as expressed through the disorder parameter $\sigma^2$ and on whether $a < c$ or $a > c$, suggested earlier in the literature [6].

Reference:
1. R.M. Kusters, J. Singleton, D.A. Keen, R. McGreevy and W. Hayes, Physica B **155**, 362 (1989)
2. R. Von Helmholt, J. Wecker, B. Holzapfel, L. Schultz and K. Samwer, Phys. Rev. Lett. **71**, 2331 (1993)
3. K. Chahara, T. Ohno, M. Kasai and Y. Kozono, Appl. Phys. Lett. **63**, 1993 (1993)
4. S. Jin, T.H. Tiffel, M. McCormack, R.A. Fastnacht, R. Ramesh and L.H. Chen, Science **264**, 413 (1994)
5. A. Urushibara, Y. Moritomo, T. Arima, A. Asamitsu, G. Kido and Y. Tokura, Phys. Rev. B **51,** 14103 (1995)
6. L. M. Rodriguez-Martinez and J. P. Attfield, Phys. Rev. B **63**, 024424-1 (2000)
7. L. M. Rodriguez-Martinez and J. P. Attfield, Phys. Rev. B **54**, R15622 (1996)
8. I. S. Jacobs, J. Phys. Chem. Solids **11**, 1 (1959)
9. W. Abdul-Razzaq and Min Wu, Superlattices and Microstructures **29**, 273 (2001)
10. P.G. de Gennes, Phys. Rev. **118**, 141 (1960)
11. Rakesh Kumar et al Physica B
12. D. Akahoshi, M. Uchida, Y. Tomioka, T. Arima, Y. Matsui and Y. Tokura, Phys. Rev. Lett. **90**, 177203 (2003)

**Figure Captions**

**Fig. 1.** Magnetization as a function of temperature for the compounds $La_{0.8-x}Eu_xSr_{0.2}MnO_3$ ($0.04 \leq x \leq 0.36$) in a magnetic field of 50 Oe. Inset shows magnetization as a function of magnetic field at 5 K.

**Fig. 2.** Magnetic ordering temperature $T_C$ and metal-insulator transition temperatures as a function of $x$ for $La_{0.8-x}Eu_xSr_{0.2}MnO3$ ($0.04 \leq x \leq 0.36$). The change in variance as a function of $x$ is also shown.

**Fig. 3.** AC susceptibility as a function of temperature for the compounds $La_{0.8-x}Eu_xSr_{0.2}MnO_3$.

**Fig. 4.** Resistivity as a function of temperature for the compounds $La_{0.8-x}Eu_xSr_{0.2}MnO_3$ ($0.04 \leq x \leq 0.36$) in a magnetic field of 0 kOe and 90 kOe. Right panel shows the magnetoresistance corresponding to a magnetic field of 90 kOe.

**Fig. 1**

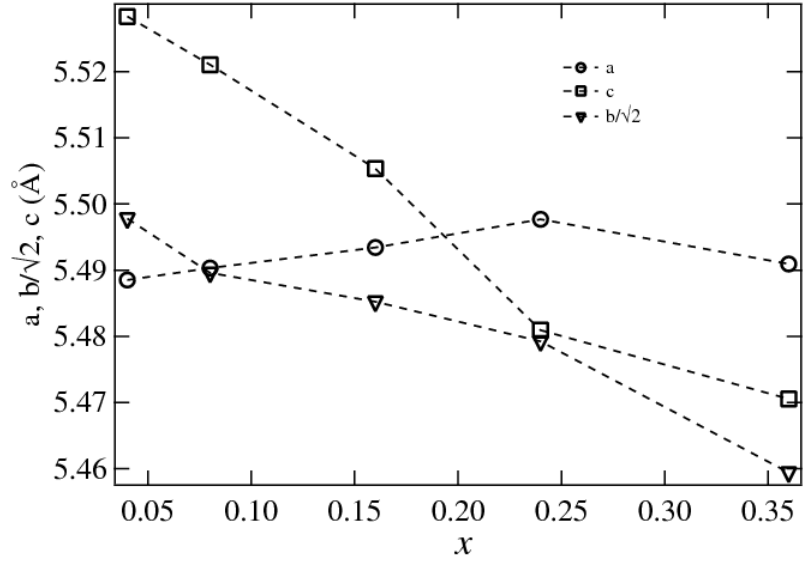

**Fig. 2**

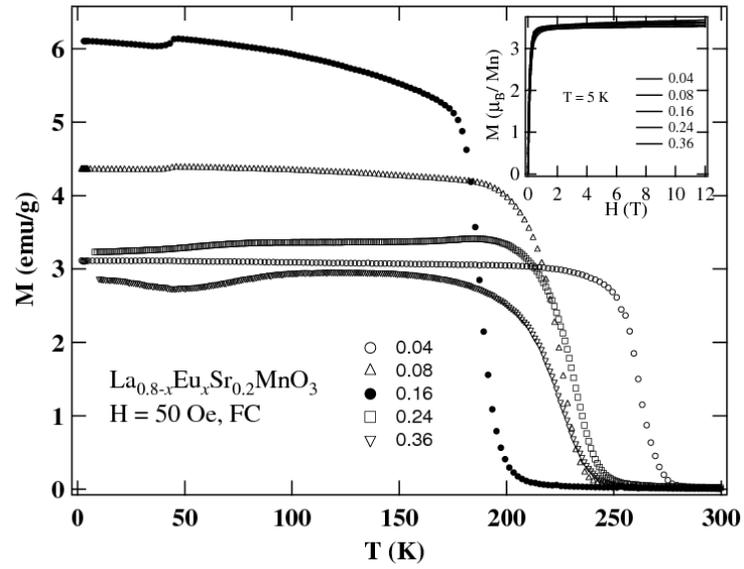

**Fig. 3**

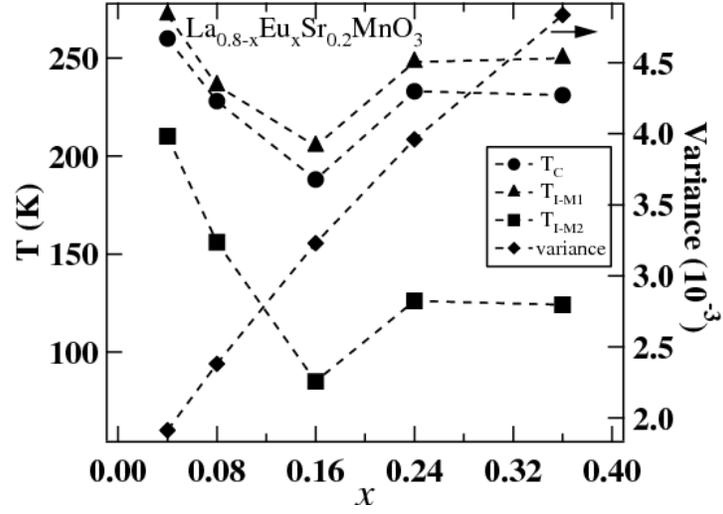

Fig. 4

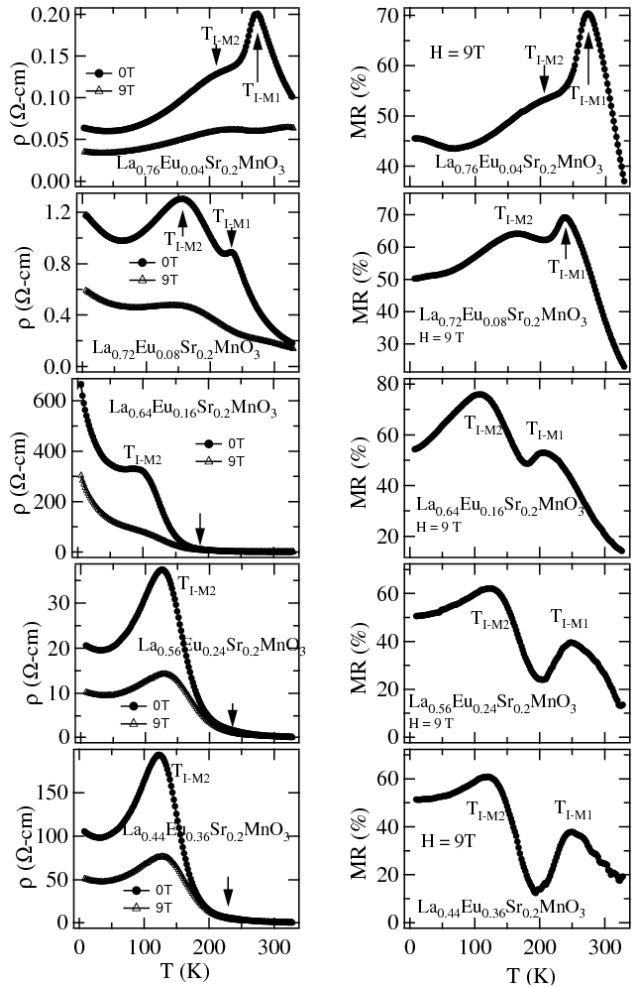